\documentclass[sigconf,nonacm]{acmart}
\usepackage{listings}

\AtBeginDocument{%
  \providecommand\BibTeX{{%
    \normalfont B\kern-0.5em{\scshape i\kern-0.25em b}\kern-0.8em\TeX}}}

\usepackage{url}
\usepackage{enumitem}
\usepackage{color}

\begin{document}

\title{PyPoll: A python library automating mining of networks, discussions and polarization on Twitter}

\author{Dimitrios P. Giakatos}
\email{dgiakatos@csd.auth.gr}
\affiliation{%
  \institution{Aristotle University of Thessaloniki}
  \country{Greece}
}

\author{Pavlos Sermpezis}
\email{sermpezis@csd.auth.gr}
\affiliation{%
  \institution{Aristotle University of Thessaloniki}
  \country{Greece}
}

\author{Athena Vakali}
\email{avakali@csd.auth.gr}
\affiliation{%
  \institution{Aristotle University of Thessaloniki}
  \country{Greece}
}

\renewcommand{\shortauthors}{D.P. Giakatos et al.}

\begin{abstract}
Today online social networks have a high impact in our society as more and more people use them for communicating with each other, express their opinions, participating in public discussions, etc. In particular, Twitter is one of the most popular social network platforms people mainly use for political discussions. This attracted the interest of many research studies that analyzed social phenomena on Twitter, by collecting data, analysing communication patterns, and exploring the structure of user networks. While previous works share many common methodologies for data collection and analysis, these are mainly re-implemented every time by researchers in a custom way. In this paper, we introduce PyPoll an open-source Python library that operationalizes common analysis tasks for Twitter discussions. With PyPoll users can perform Twitter graph mining, calculate the polarization index and generate interactive visualizations without needing third-party tools. We believe that PyPoll can help researchers automate their tasks by giving them methods that are easy to use. Also, we demonstrate the use of the library by presenting two use cases; the PyPoll visualization app, an online application for graph visualizing and sharing, and the Political Lighthouse, a Web portal for displaying the polarization in various political topics on Twitter. 
\end{abstract}

\ccsdesc[500]{Networks~Online social networks}
\ccsdesc[500]{Software and its engineering~Software libraries and repositories}
\ccsdesc[300]{Human-centered computing~Visualization toolkits}

\keywords{Online social networks, graph mining, polarization, open-source}



\maketitle

\section{Introduction}
Nowadays, online social networks/media play a vital role in public discussions of various topics all around the world \cite{matakos2017measuring,kubin2021role,zhuravskaya2020political,valenzuela2012social,stieglitz2013social}. In 2022, 190 million new users entered the world of social media, meaning they started to read news and participate in public discussions about many topics \citep{datareportal2023globalsocialmediastatistics}. Twitter is one of the largest social media platforms where users are free to express their opinion within 280 characters of text ("tweets") and engage in public debates. Because of the text length limit, Twitter is used, in particular, by many politicians for posting short statements about political, economic, or social topics to inform the public. On the other hand, citizens themselves use to post their own statements, and discuss with others by responding to ("quote") or sharing ("retweet") others' posts. As a result, Twitter plays a significant role in modern political discussions and, frequently, opinion formation.

\textbf{Graph mining and analysis on Twitter:} User interactions on Twitter (e.g., through comments, retweets, mentions, etc.) can be modeled as a graph, where nodes represent users and edges represent interactions between them. Graphs and graph analyses have been used to model and explore various phenomena on social media, such as recommendation systems~\cite{schroeder2019fact} (e.g, predict new connections between tweets, hashtags, and users)
, bot detection~\cite{alothali2018detecting} (e.g., find if a user is a bot according to his followers), or topic sentiment analysis~\cite{wang2011topic} (i.e. connect hashtags that co-occur in Tweets and use node features and graph structure for node classification). 

\textbf{Polarization on Twitter:} In particular, an important phenomenon is \textit{polarization}, where users participating in an online discussion start to engage only with like-minded users and/or adopt ideas of extreme views \cite{matakos2017measuring,isenberg1986group}. While this phenomenon exists in every society~\cite{isenberg1986group}, it is frequently amplified in the online world \cite{iandoli2021impact}. A social network with intense polarization comprises two separate groups of users with confronting opinions. In a polarized group, if someone starts to share an opinion that differs from the opinion that the group accepts, then he has a high percentage of being rejected by the community. As a result, the users of both groups tend to adopt a single perspective, thus leading to a non democratic discussion.

\textbf{One common research methodology, but lack of common tools:} What is common in the majority of research methodologies and previous works on Twitter analytics is the graph mining pipeline that consists of two parts: (i) A set of tweets are collected using the Twitter API. The tweets are typically related to a specific discussion, which is done by using one or more keywords or hashtags in the API call. (ii) The set of users and their interactions are extracted from the meta-data and other fields included in the collected tweets, and a graph is then created. Moreover, in the case of polarization, several research methods need to annotate the users in order to assign them in one of the opinion groups; while, here, we can observe more variations in the research methods (e.g., based on the content, such as usage of common hashtags~\cite{matakos2017measuring}, tweets~\cite{garimella2018quantifying}, or retweets~\cite{garimella2018quantifying}), in many works the users are assigned to a group based on the accounts they follow. For example, followers of different politicians can be attributed to different groups~\cite{matakos2017measuring,garimella2018quantifying}. 


Despite the fact that many works share this common methodology, to our best knowledge, there are no available open-source tools for operationalizing the graph creation from a Twitter discussion (neither for automatic the polarization analysis). Existing solutions include scripts that can create Twitter graphs based on the followers~\citep{github2012twittersocialgraphnetworkx,github2016twitterkeywordgraph,github2018twittergraph,github2023twittergraph}, but without tracking a discussion and capturing it in a graph. 
As a result, researchers and practitioners need to develop from scratch their methods, which requires significant effort or can even be a serious impediment for non-technical users (e.g., from domains other than computer science that are interested in the social aspects of Twitter discussions).

\begin{figure}
  \includegraphics[width=\linewidth]{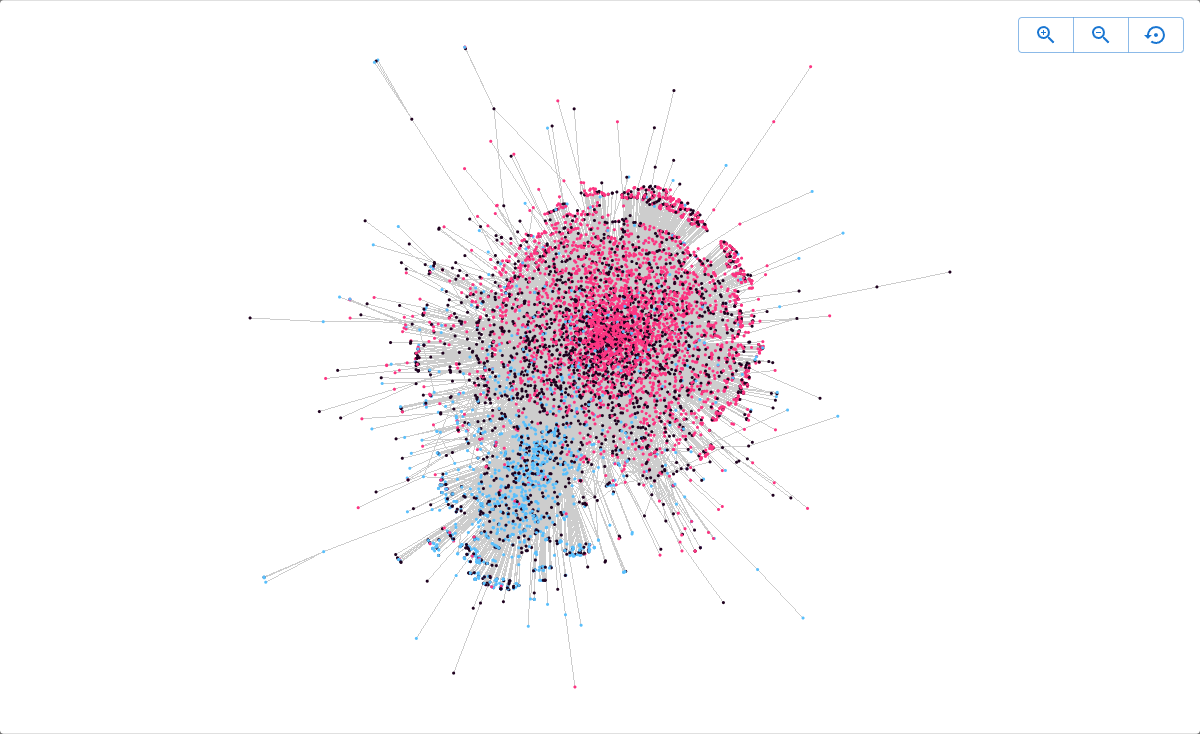}
  \caption{Graph visualization using PyPoll}
  \label{fig:graph_visualization}
\end{figure}

\textbf{Contributions}: Motivated by this lack of common tools for these common methodologies, in this paper, we introduce \textit{PyPoll}\footnote{PyPoll: \textbf{Py}thon Twitter mining and \textbf{pol}arization \textbf{l}ibrary \cite{PyPoll}.} (Section~\ref{sec:PyPoll}), an open-source Python library that automates the mining, graph creation, and polarization analysis from Twitter discussions. Specifically, using PyPoll a user can automate the following tasks:

\begin{itemize}[leftmargin=*]
    \item \textit{Twitter graph mining:} Using as input a list of keywords/hashtags and a single method (one line of code), PyPoll collects all tweets relevant to a discussion, saves them to a database, and builds a graph that includes also node and edge features.
    \item \textit{Polarization:} PyPoll can easily collect information about followers of Twitter users, and based on this to annotate the nodes (i.e., users) of the constructed graph. Moreover, it contains methods that use this information to calculate the polarization index of the discussion.
    \item \textit{Visualization:} Finally, PyPoll can create graph layouts and visualize graphs without using any third-part tools (see Fig.\ref{fig:graph_visualization}).
\end{itemize}

PyPoll is open-source and guidelines and examples for using it can be found on GitHub \cite{PyPoll}. Furthermore, we demonstrate its use through two example cases (Section~\ref{sec:use-cases}): (i) a visualization Web application where the user can upload a graph file generated by PyPoll (e.g., from their own collected dataset) and produce a shareable visualization of the Twitter graph, annotated with user group information, and (ii) a Web portal where we present data and visualizations for several political discussions we analyzed.

\section{The PyPoll library}\label{sec:PyPoll}
\begin{figure*}
  \includegraphics[width=\textwidth]{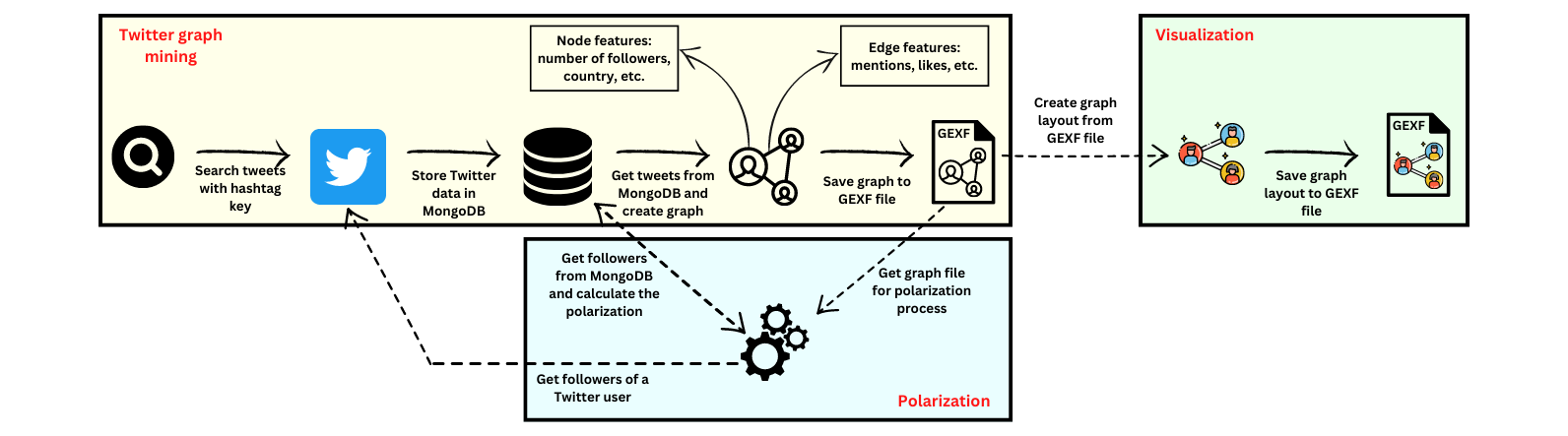}
  \caption{PyPoll: Polarization Library pipeline.}
  \label{fig:PolL}
\end{figure*}

The overall architecture of the library is depicted in Fig.~\ref{fig:PolL}, and, in the following, we present the design and functionality of its three components: Twitter graph mining, polarization analysis, and visualization.

\subsection{Twitter graph mining}\label{sec:graph-minining}

\noindent\textbf{Input:} A list of keywords and/or hashtags.

\noindent\textbf{Output:} A graph with node and edge attributes.

\noindent\textbf{Overview:} PyPoll receives as input a list of keywords and/or hashtags relevant to a discussion in Twitter, calls the Twitter API, collects all the returned tweets and stores them in a database. Then, it extracts information contained in the collected tweets to build a graph, where nodes represent Twitter users and edges interactions between them, and enriches the graph with node attributes extracted from the tweets (e.g., nb of tweets, followers, username, i.e.) and edge attributes (namely, number and type -retweet, mention, comment- of interactions). 

While this Twitter graph mining process contains several steps with technical challenges (as described below), PyPoll automates it and enables users to mine and construct a graph with a few lines of code as seen in the example of Listing~\ref{code1}.

\textit{Collecting tweets:} PyPoll wraps the tweepy library \citep{github2023tweepy} for collecting data from the Twitter API. In practice, several problems may happen during a call to the Twitter API (e.g., with a list of keywords) that will break the collection process. PyPoll handles errors that may occur, and automatically resumes the collection from where it stopped. Moreover, in the case a user wants to update an existing collection of tweets, PyPoll automatically detects the last saved tweet based on its creation date and time, and collects only the missing (newest) content. Finally, the collected tweets are stored to a local database (MongoDB). All this functionality can be accessed through a single method \texttt{get\_tweets} (see line~4 in Listing~\ref{code1}).

\textit{Constructing the graph:} The construction of a graph from the collected data is again accessed through the method \texttt{create\_graph}. When calling this method, PyPoll extracts from the tweets in the collection: (i) the Twitter users, representing them as nodes, (ii) attributes of the users contained in the tweets' meta data (e.g., username) or calculated on the collection (e.g., number of tweets in the discussion), (iii) the interactions between users from the tweets' metadata, representing them as edges and edge attributes (e.g., a user retweeting two tweets of another user is represented as an edge of type "retweet" and weight 2). This functionality allows the user to create different graphs (from the same collection of tweets) based on optional arguments passed to the library. For example, depending on the use case, a researcher may need to consider only retweets or only likes as graph edges \citep{bild2015aggregate, yang2012finding,song2011spam} (e.g. paper \citep{bild2015aggregate} used retweets for spam detection).

Finally, after creating the graph, PyPoll saves it in a format selected by the user (currently available options are the -popular- formats: GEXF, GML, JSON).

\begin{lstlisting}[language=Python,frame=single,caption=Example script for Twitter graph mining,label=code1]
from pypoll import Twitter, MongoDB, Graph
db = MongoDB(...)
api = Twitter(twitter_bearer_token, db)
api.get_tweets("#test")
graph = Graph()
graph.create_graph("#test", db, {...})
\end{lstlisting}

\subsection{Polarization}\label{sec:polarization}

\noindent\textbf{Input:} Two or more Twitter accounts.

\noindent\textbf{Output:} Polarization analysis of the graph.

\noindent\textbf{Overview:} Given a set of Twitter accounts (corresponding to different "opinions", e.g., accounts of two political parties), PyPoll collects their followers, finds which of them are also in an existing collection of tweets (Section~\ref{sec:graph-minining}), annotates the nodes in the Twitter graph, and calculates the polarization index of the graph. 

These polarization analysis steps are automated by PyPoll, and can be accessed through a few lines of code as in the example of Listing~\ref{code2}.


\textit{Opinion attribution:} In the polarization analysis we consider (and similarly to several previous works \citep{matakos2017measuring,garimella2018quantifying}), users are assigned to an "opinion" (or, "group" or community) based on the account(s) they follow. For example, following the account of a specific political party can be used as an indication for the political attribution of the user \citep{matakos2017measuring}. PyPoll implements this methodology, by collecting all the followers of two given Twitter accounts (see method \texttt{get\_followers} in lines 1-2 in Listing~\ref{code2}). Similarly to the graph mining methods (Section~\ref{sec:graph-minining}), PyPoll handles in the background potential errors in the follower collection process. Then, it finds which of the collected followers appear also in an existing collection of tweets, and adds the inferred attributed opinion as an extra node label for these users.

\textit{Polarization index:} After the node labeling, the user can use PyPoll to calculate the polarization of the graph/discussion, using the \texttt{get\_polarization} method (line 4 in Listing~\ref{code2}), which takes as input a polarization metric, such as the Friedkin \& Johnsen \cite{matakos2017measuring} or the Random Walk Controversy \cite{garimella2018quantifying}. PyPoll implements the corresponding methods and calculates the metric, which takes values in the range between 0 (no polarization) to 1 (very polarized discussion).

\begin{lstlisting}[language=Python,frame=single,caption=Example code for polarization analysis of a graph,label=code2]
...
api.get_followers("username_A")
api.get_followers("username_B")
...
options = {
    ...
    "users": {"username_A", "username_B"}
}
graph.create_graph("#test", db, options)
graph.get_polarization(["fj", "rwc"])
\end{lstlisting}

\subsection{Visualization}
\noindent\textbf{Input:} A Twitter graph file.

\noindent\textbf{Output:} An interactive visualization of the graph.

\noindent\textbf{Overview:} Visualization of graphs can sometimes quickly reveal insights that are not easily captured by graph metrics, e.g., structure of communities, and thus is a useful component several graph analyses tasks. PyPoll can be used to easily (see example code in Listing~\ref{code3}) create a graph layout and visualize a generated graph (Section~\ref{sec:graph-minining}), with the option of coloring nodes according to their attributed opinions (Section~\ref{sec:polarization}). An example visualization of PyPoll is given in Fig.~\ref{fig:graph_visualization}.

PyPoll implements the Fruchterman-Reingold force-directed algorithm \citep{fruchterman1991graph} to create the graph layout, as well as an in-build visualization tool that enables the user to interact with the graph (e.g, zoom-in/out, navigation, node selection). It is important to note that with PyPoll, the user does not need any third-party visualization tools.



%

\begin{lstlisting}[language=Python,frame=single,caption=Example code for graph visualization,label=code3]
...
graph.create_layout("#test.gexf", {...})
plot = GraphPlot()
plot.show("#test.gexf")
\end{lstlisting}



\section{Use Cases}\label{sec:use-cases}

We demonstrate the use of PyPoll through two use cases, a visualization Web application (Section~\ref{sec:webapp}) and a Web portal with polarization analyses of political discussions on Twitter (Section~\ref{sec:lighthouse}).

\subsection{PyPoll visualization app}\label{sec:webapp}

We implemented a Web application, where users can easily create and share online visualization of graphs. A user uploads a GEXF graph file (e.g.,  created by PyPoll), and receives a URL with an online visualization of the uploaded graph. The URL can be shared with anyone and other users can view and interact with the graph online through the browser without restrictions (i.e. create an account). For security reasons, users that want to create visualizations need to create an account (for free) and login, whereas for accessing a visualization no account is needed.\footnote{\url{https://dpgiakatos.com/redirections/pypoll-visualization-app}}



The Web application consists of three components: the frontend, the backend and the Python SDK. The frontend is implemented in ReactJS\footnote{\url{https://reactjs.org}} (HTML/CSS/JavaScript) and deployed in a NGINX web server\footnote{\url{https://www.nginx.com}}. The backend is implemented with FastAPI\footnote{\url{https://fastapi.tiangolo.com}} and MongoDB and the Python SDK is implemented in Python. The code for the entire Web app is available on GitHub\footnote{\url{https://github.com/dpgiakatos/PyPoll-app}} under the MIT license. 

\subsection{Political Lighthouse}\label{sec:lighthouse}
A use case for PyPoll is the exploration of discussions on Twitter related to political topics. To this end, we developed the Political Lighthouse\footnote{\url{https://dpgiakatos.com/redirections/political-lighthouse}}, a web page that explores polarization in Greek political discussions on Twitter. 

Specifically, we collected tweets for several political discussions, including topics related to political parties, the energy crisis, and national issues. Each discussion corresponds to a different hashtag that was trending during the collection period. For the opinion attribution, we collected the followers of the accounts corresponding to heads of two major political parties (a liberal-conservative and a left-wing political party). Then, we visualized the generated graphs and presented them in the web page, along with statistics about the discussions (number of participating users, number of tweets, etc.) and their polarization. 


\section{Conclusion}
In this paper we have presented PyPoll, an open-source Python library that can be used to automate collection of tweets relevant to a discussion, create a graph out of them, collect data and analyze polarization, and visualize graphs. Our design principles for PyPoll were generality (e.g., can be applied to any discussion, and offers several options for graph creation and visualization) and ease of use. Researchers and practitioners can use PyPoll to create, analyze and visualize Twitter discussions with few lines of codes, thus avoiding technical complexities involved in these tasks. Furthermore, through the PyPoll visualization app we demonstrated, users can share their graphs about different topics easily. We believe that PyPoll can facilitate research in social computing topics; as in the example of the Political Lighthouse web page we presented. 

PyPoll can be further extended in the future, e.g., by implementing more opinion attribution methods (e.g., based on content, such as, hashtags used in tweets), or more polarization metrics and methodologies.

\begin{acks}
This work was co-financed by the European Union and Greek national funds through the Operational Program for Competitiveness, Entrepreneurship and Innovation, under the call RESEARCH-CREATE-INNOVATE (Project Code: T2EDK-04937).
\end{acks}
\bibliographystyle{ACM-Reference-Format}
\bibliography{references}

\end{document}